\documentclass{article}

\usepackage{amsmath, amscd, amssymb, mathrsfs, tikz-cd, verbatim, parskip}
\usepackage[hang,flushmargin]{footmisc}
\usepackage[a4paper, total={6in, 9in}]{geometry}
\newcommand{\pa} [2] {\frac{\partial #1}{\partial #2}} 

\begin{document}

\begin{center}
\Huge On the Kostant-Souriau prequantization of scalar fields with polysymplectic structures

\vspace{5mm}

\Large Tom McClain \footnote{mcclaint@wlu.edu, ORCiD 0000-0002-2330-3501 }

Department of Physics and Engineering 

Washington and Lee University, Lexington VA 24450

\vspace{1cm}

\normalsize

\begin{abstract}
In this paper, I present a novel, purely differential geometric approach to the quantization of scalar fields, with a special focus on the familiar case of Minkowski spacetimes. This approach is based on using the natural geometric structures of polysymplectic Hamiltonian field theory to produce an analog of the Kostant-Souriau prequantization map familiar from geometric quantization. I show that while the resulting operators are quite different from those of canonical quantum field theory, the approach is nonetheless able to reproduce a few of canonical quantum field theory's most fundamental results. I finish by elaborating the current limitations of this approach and briefly discussing future prospects.
\end{abstract}

Pacs numbers: 02.40.-k, 03.70.+k


\end{center}

\section{Introduction} \label{intro}

From the beginning, one of the main goals of the polysymplectic and multisymplectic approaches to Hamiltonian field theory has been to provide a differential geometric setting for classical field theories that would make them easier to quantize using non-canonical methods of quantization -- in particular geometric quantization \cite{kijowski1976canonical} \cite{gotay1991multisymplectic} -- or at least to ease the challenge of understanding the structure of the (quantum field) theories that result from their quantization \cite{gunther1987polysymplectic}. A great deal of progress has been made in developing the mathematics of non-canonical quantization for field theories since the pioneering work in these fields, especially from the perspective of deformation quantization; see, for example, the recent works of Fredenhagen and Rejznek \cite{fredenhagen2015perturbative} and Berra-Montiel and Molgado \cite{berra2020coherent}. Yet relatively little has been done to develop purely differential geometric quantization methods that start from the polysymplectic or multisymplectic perspective, nor to relate these mathematical ideas to the experimentally verifiable physics of canonical quantum field theory; the works of Kanatchikov \cite{kanatchikov1998toward}, Navarro \cite{navarro1998toward}, and von Hippel and Wohlfarth \cite{von2006covariant}, are a few exceptions that I will later directly compare to the approach I articulate here.

The goal of this work is twofold. First, I present a novel, purely differential geometric approach to the quantization of classical fields based on the natural structures of polysymplectic field theory used in \cite{mcclain2021global}. Second, I apply this quantization method to the simplest physically reasonable field theory, namely that of a single real-valued scalar field in flat (Minkowski) spacetime. Though this approach seems too simplistic to succeed, I show that it nonetheless reproduces a few of the key concrete features of quantized scalar fields in canonical quantum field theory.

Before launching into the details, a brief overview is in order. In section \ref{polysymplecticfieldtheory} I outline my favored formulation of the polysymplectic approach to \emph{classical} field theory; even for those readers already familiar with polysymplectic field theory, this section is necessary to establish my notation and conventions. In section \ref{ksqqpss} I outline my novel approach to prequantization with polysymplectic structures. In section \ref{qftfrompq} I point out the inability of this quantization process to specify explicit spacetime dependence for the resulting operators, then show how these operators are nonetheless ultimately able to reproduce an integrated version of the canonical commutation relations among the extended phase space coordinate functions, as well as the proper eigenvalue equations for single particle energy-momentum eigenstates. I discuss some of the merits and limitations of this approach in section \ref{discussion}, concluding with a few remarks about applications and directions for future work in section \ref{conclusions}.

\section{Polysymplectic Hamiltonian field theory} \label{polysymplecticfieldtheory}

To get us to my proposed quantum field theoretic framework, it is necessary to take a rather large detour into polysymplectic Hamiltonian field theory, which will ultimately make it possible to use a generalization of the Kostant-Souriau prequantization map on classical field theories.

Pioneered by Christian G\"unther in the 1980s \cite{gunther1987polysymplectic}, the polysymplectic approach to Hamiltonian field theory is -- along with the original de Donder-Weyl and multisymplectic \cite{gotay1991multisymplectic} approaches -- a covariant alternative to the usual canonical approach to Hamiltonian field theory. In the polysymplectic approach, the canonical conjugate momentum field(s) is replaced by a larger number of polymomenta, the symplectic structure is a vector-valued two-form, and the geometric setting is a finite-dimensional manifold rather than the (infinite-dimensional) space of solutions. My recent work has extended G\"unther's analysis from the local/flat setting to curved manifolds \cite{mcclain2021global}. As a simple example to fix our ideas and establish notational conventions, let us briefly examine the case of a single, real-valued scalar field $\phi$ in a general curved spacetime $M$ with metric $g$. All the details can be found in \cite{mcclain2021global}.

If the classical configuration space for the field is $E = M \otimes \mathbb{R}$, the geometric setting for the polysymplectic theory is the extended phase space $P = V^*E \otimes TM$, where $VE$ denotes the vertical bundle of the fiber bundle $E$ and $V^*E$ is the fiber-wise dual space to $VE$. Local fibered coordinates on $P$ look like $\{ x^\mu, \phi, \pi^\mu \}$, where the spacetime index $\mu$ runs over the values $\{ 0, 1, 2, 3 \}$. In appropriate local coordinates, the polysymplectic form $\omega$ looks like

\begin{equation} \label{polysymplectic} \omega = d \phi \wedge d \pi^\mu \otimes \pa{}{x^\mu} \end{equation}

where the $\wedge$ represents the exterior product. Hamilton's field equations for solutions $\gamma : M \to P$ come from the requirement that, for a suitable vertical projection $V_P: TP \to VP$ \footnote{In the case of the Klein-Gordon field, this vertical projection is a natural result of using the Levi-Civita connection and ordinary exterior derivative to construct $V_P$; other situations sometimes require more care.} and for all vertical vector fields $v : M \to VP$, we have

\begin{equation} \label{hamiltons} \omega(V_P \circ T \gamma, v) = dH(v) \end{equation}

where $T\gamma : M \to TP$ is the tangent map for the particular solution $\gamma$ and $H : P \to \mathbb{R}$ is an appropriate function called the covariant Hamiltonian function. (Note that the covariant Hamiltonian function is generally not the same as the Hamiltonian density from the canonical formalism.)

In this particular case, we naturally recover the (curved spacetime) Klein-Gordon equation

\begin{equation} \label{kleingordonequation} g^{\mu \nu} \nabla_\mu \nabla_\nu \phi + m^2 \phi = 0 \end{equation} 

where $\nabla$ represents the covariant derivative with respect to the Levi-Civita connection associated with the metric $g$.

Critically, the polysymplectic approach also gives us a kind of generalized Poisson tensor $\Pi$, represented in appropriate local fibered coordinates $\{ x^\mu, \phi, \pi^\mu \}$ as

\begin{equation} \label{poisson} \Pi = - \pa{}{\phi} \wedge \pa{}{\pi^\mu} \otimes dx^\mu \end{equation}

and a vector-valued polysymplectic potential $\theta$, represented in the same coordinates as

\begin{equation} \label{potential} \theta = \pi^\mu ( d \phi + V_\nu d x^\nu) \otimes \pa{}{x^\mu} \end{equation}

where the $V_\nu$ are the components of an appropriate vertical projection on $VE$ that can be derived from the vertical projection $V_P$ cited above.\footnote{In the case of the real scalar field, these components are all zero.} $\Pi$ and $\theta$ are the natural geometric structures I will use to construct my prequantization map.

\section{Kostant-Souriau prequantization with polysymplectic structures} \label{ksqqpss}

From the perspective of quantization, the main benefit of the polysymplectic approach is that it mimics very closely the geometric structures of Hamiltonian particle mechanics that form the basis for most successful differential geometric quantization schemes. That being the case, there are several well-established approaches to quantization that it makes sense to try to emulate with polysymplectic structures:

\begin{enumerate}
\item Kostant-Souriau prequantization with the generalized Poisson tensor and polysymplectic potential 
\item Geometric quantization with a connection whose exterior derivative relates to the polysymplectic form
\item Deformation quantization with the generalized Poisson tensor
\end{enumerate}

The simplest of these approaches is the most definitely the first, so that is the case I will analyze here. Applications of similar techniques to the other two approaches are logical subjects for future work.

The polysymplectic structures of \eqref{poisson} and \eqref{potential} can be used to produce something very much like a Kostant-Souriau prequantization map for the theory when combined with a (for now) arbitrary vector field $v : M \to TM$, one-form $\alpha : M \to T^*M$, and complex-valued function $\kappa : P \to \mathbb{C}$:

\begin{multline} \label{abstractquantization}
Q_{v, \alpha, \kappa} : f \to L(C^1(M) \to C^1(M)) \mid f \mapsto f + \Pi(\alpha(\theta), df, v) + \kappa \Pi(-, df, v) \\
= f - \alpha_\nu \pi^\nu v^\mu \pa{f}{\pi^\mu} + \kappa v^\mu \left(\pa{f}{\phi} \pa{}{\pi^\mu} - \pa{f}{\pi^\mu} \pa{}{\phi} \right) \end{multline}

This map takes differentiable functions on the extended phase space $P$ considered in section \ref{polysymplecticfieldtheory} and turns them into linear operators on the space of differentiable functions over that same extended phase space. Though it is defined primarily in terms of geometric structures natural to the polysymplectic theory, it depends critically on the arbitrary vector field $v$, one-form $\alpha$, and function $\kappa$.

Note that so far this map is very much like the ones introduced almost simultaneously by Kanatchikov in \cite{kanatchikov1998toward} and Navarro in \cite{navarro1998toward}. However, due to the explicit use of the polysympectic structures of \cite{mcclain2021global}, \eqref{abstractquantization} is produced very naturally here by the most direct possible analogy with the simplest Kostant-Souriau prequantization procedure for ordinary Hamiltonian particle theory. This is in contrast to being built up from more complicated mathematical manipulations of the polymomentum phase space as in \cite{kanatchikov1998toward} or being posited (rather than derived) by appeal to more general physical principles as in \cite{navarro1998toward}. In fact, the approach that seems the most closely related is the one used later by von Hippel and Wohlfarth in \cite{von2006covariant}, though there the authors are careful to avoid explicit reference to any global structures whatsoever. Perhaps more importantly, in all three of these similar cases the authors chose to use Dirac gamma matrices (or some generalization) in place of the vector field $v$ I use above\footnote{This despite the fact that von Hippel and Wohlfarth explicitly mention and reject the possibility of using a vector field rather than the gamma matrices in \cite{von2006covariant}.}. This is all to say that though the above construction is superficially very similar to these earlier approaches, it is in fact novel.

To make contact with something more physical, let us assume that $v$ is a non-vanishing, timelike vector field and that $\alpha$ is its dual, normalized so that $\alpha(v) = 1$. (This is always possible in a globally hyperbolic spacetime of the type usually considered in quantum field theory on curved spacetime \cite{hollands2015quantum}.) Dimensional analysis suggests that the units of $\kappa$ must be $[ \kappa ] = [ \pi^\mu ]^2$. For the standard Klein-Gordon Lagrangian density $\mathcal{L} = \frac{1}{2} g^{\mu \nu} \pa{\phi}{x^\mu} \pa{\phi}{x^\mu} - \frac{m^2 c^2}{2 \hbar^2} \phi^2$, we have $[ \pi^\mu ]^2 = \frac{M}{L T^2}$. Assuming $\kappa \propto \text{i} \hbar c$, this gives $\kappa = \frac{\text{i} \hbar c}{V_4}$, where $V_4$ is some appropriate term -- which we might naively expect to be constant -- with dimension $L^4$ on $M$. Putting this into (\ref{abstractquantization}) in a local coordinate system in which $v = v^0 \pa{}{x^0}$ gives

$$  Q_v(f) = f - \pi^0 \pa{f}{\pi^0} + \text{i} \hbar c \frac{v^0}{V_4} \left(\pa{f}{\phi} \pa{}{\pi^0} - \pa{f}{\pi^0} \pa{}{\phi} \right) $$

If we then choose to separate $\frac{1}{L^4}$ as $\frac{k}{L^3}$, we can make the convenient choice $k = \alpha_0$ (the only non-zero component of $\alpha$ in this coordinate system). Recalling that $\alpha(v) = \alpha_0 v^0 = 1$, our equation then simplifies to

\begin{equation} \label{quantization} 
Q_v(f) = f - \pi^0 \pa{f}{\pi^0} + \frac{\text{i} \hbar c}{L^3} \left(\pa{f}{\phi} \pa{}{\pi^0} - \pa{f}{\pi^0} \pa{}{\phi} \right)
\end{equation}

In these coordinates, we therefore have

$$ Q(\phi) = \phi + \frac{\text{i} \hbar c}{L^3} \pa{}{\pi^0} $$
$$ Q(\pi^0) = - \frac{\text{i} \hbar c}{L^3} \pa{}{\phi} $$
$$Q(\pi^i) = \pi^i $$

I have said that these operators act on complex-valued functions on the extended phase space $P$, but this space is too big. To find a more appropriate space of quantum states upon which our operators can act requires that we pare down this space in some appropriate way, just as in the geometric quantization of particle systems \cite{ashtekar1986geometric} \cite{woodhouse1992geometric}. One way to do this with the same polysymplectic structures used to define the prequantization map (\ref{quantization}) is to define the space $\mathbb{S}$ of quantum states via

\begin{equation} \label{states} \mathbb{S} = \{ \psi \in C^\infty(P, \mathbb{C}) \mid \Pi(d \psi, \beta(\theta), v) = 0 \ \forall \ \beta : M \to T^*M \} \end{equation}

(Note that the ingredients used here to pare down the space of quantum states are almost the same as in the middle term in (\ref{abstractquantization}), except that we have restricted our attention to the case where $v = v^0 \pa{}{x^0}$ and the one-form $\beta$ is arbitrary.)

In the same local coordinates we have been using, this amounts to the requirement that $\psi = \psi(x^\mu, \phi, \pi^i)$; i.e., wave functions cannot depend on the $\pi^0$ polymomentum coordinate.

Since differential operators of the form $O = f(x^\mu, \phi, \pi^\mu) \pa{}{\pi^0}$ are now (by construction) nilpotent on elements of our quantum state space $\mathbb{S}$, we can rewrite the quantized versions of our extended phase space coordinate functions $\{ \phi, \pi^0, \pi^i \}$ as

\begin{equation} \label{qphi} Q(\phi) = \phi \end{equation}
\begin{equation} \label{qpi0} Q(\pi^0) = - \frac{\text{i} \hbar c}{L^3} \pa{}{\phi} \end{equation}
\begin{equation} \label{qpii} Q(\pi^i) =\pi^i \end{equation}

\section{Quantum field theory from polysymplectic quantization?} \label{qftfrompq}

The foregoing is all rather elegant, quickly and easily quantizing the coordinate functions on the extended phase space $P$ while sidestepping almost every ordinary mathematical challenge associated with the quantization of classical fields. However, it is such an alien construction compared to canonical quantum field theory that is difficult to see how to compare the two.

Note, for example, the lack of explicit spacetime dependence in the operators of \eqref{qphi}, \eqref{qpi0}, and \eqref{qpii}; this is a natural consequence of directly quantizing the extended phase space coordinate functions, rather than working with the space of solutions to the classical Klein-Gordon equation as in the canonical approach. Looking back at (\ref{hamiltons}) makes it clear how spacetime dependence is incorporated in classical polysymplectic field theory: just as in geometric Hamiltonian particle mechanics, we make contact between our polysymplectic structures and actual solution sections only by pulling back our geometric structures by prospective solution sections ``at the last minute." 

Indeed, it is possible to give spacetime dependence to the operators of \eqref{qphi} and \eqref{qpii} by pullback by classical solutions to \eqref{kleingordonequation}. In fact, it is even possible to give spacetime dependence to the operator of \eqref{qpi0} through a non-unique extension of this kind of approach. However, as the price of this extension seems to be a breakdown of mathematical elegance and a return to the theory of distributions, I will not pursue this approach here. Instead, we will see how far a more geometric approach can take us.

The operators of \eqref{qphi}, \eqref{qpi0}, and \eqref{qpii} all naturally commute with one another except for the commutator

\begin{equation} \label{psqccr} [ Q(\phi), Q(\pi^0)] = \frac{\text{i} \hbar c}{L^3} \end{equation}

If we integrate \eqref{psqccr} over an arbitrary spatial hypersurface of constant time, we find

$$ \int dV [Q(\phi), Q(\pi^0)] = \frac{\text{i} \hbar c \, V}{L^3} $$

where $V$ is the spatial volume of the hypersurface. Had we integrated the ordinary commutation relation between the field operator $\hat \phi(\vec x)$ and the conjugate momentum density operator $\hat \pi(\hat y)$ from canonical quantum field theory, we would have found

$$ \int dV [\hat \phi(\vec x), \hat \pi(\vec y)] = \text{i} \hbar c $$

Requiring these two to match suggests that we should take $ L^3 = V $, so that \eqref{qpi0} becomes

\begin{equation} \label{qpi0rev} Q(\pi^0) = - \frac{\text{i} \hbar c}{V} \pa{}{\phi} \end{equation}

Though taking this path allows us to match the \emph{integrated} commutation relation of canonical quantum field theory, this might not seem like a particularly promising approach. However, it turns out that it also allows us to successfully quantize the classical total field momentum $P^j =  \frac{1}{c} \int dV \, T^{0j} = \frac{1}{c} \int dV \, \pi^0 \pi^j $. 

To do so, we first quantize the extended phase space function $T^{0j} = \pi^0 \pi^j$ using \eqref{quantization} to find

$$Q(T^{0j}) = \pi^j Q(\pi^0) = - \frac{\text{i} \hbar c}{V} \pi^j \pa{}{\phi} $$

Since none of the extended phase space operators carries any explicit spacetime dependence, it is easy to integrate this operator over any spatial hypersurface to find

\begin{equation} \label{totalmomentumoperator} Q(P^j) = \frac{1}{c} \int dV \, Q(T^{0 j}) = - \text{i} \hbar \, \pi^j \pa{}{\phi} \end{equation}

If we now act this operator on a hypothetical extended phase space state $\Psi \in \mathbb{S} = \Psi(x^\mu, \phi, \pi^i)$, we find that

\begin{equation} \label{phasespacemomentumeigenfunction} Q(P^j) \Psi = - \text{i} \hbar \, \pi^j \pa{\Psi}{\phi} \end{equation}

So far, this is all quite simple and natural, but we have not found ourselves any closer to the results of canonical quantum field theory. However, if we remember that in classical  polysymplectic field theory we ultimately need to pullback our differential geometric structures by solution sections to make contact with physics, we might be motivated to pullback \eqref{phasespacemomentumeigenfunction} by a complex-valued energy-momentum eigenfunction $\gamma(x^\mu) = A e^{\text{i} g_{\mu \nu} k^\mu x^\nu}$ of the Klein-Gordon equation \eqref{kleingordonequation} in Minkowski space to find

$$\gamma^* \left( Q(P^j) \Psi \right) = - \text{i} \hbar \, \text{i} k^j \gamma \, \gamma^*(\Psi) = \hbar k^j \gamma \, \gamma^*\left(\pa{\Psi}{\phi}\right) $$

where we have assumed that our metric has signature $(+,-,-,-)$ and remembered that $\gamma^*(\pi^j) = g^{j \mu} \pa{\gamma}{x^\mu} = \text{i} k^j \gamma$.

If we now choose the specific extended phase space state $\Psi_{s}(x^\mu, \phi, \pi^i) \propto e^{\frac{1}{C^2} \phi \bar \phi}$, where $C$ is some as yet undefined constant with the same units as $\phi$ (that is, $[C] = [\phi]$), we find that

\begin{equation} \label{pullbackmomentumeigenfunction} \gamma^* \left( Q(P^j) \Psi_s \right) = \hbar k^j \frac{A \bar A}{C^2} \, \gamma^*(\Psi_s) \end{equation}

Thus, with this choice of $\Psi$ and the particular choice $C^2 = A \bar A$, we are able to match -- after pullback -- the usual result for the total momentum of a single quantum particle in the momentum eigenstate $\vec k$.

Since $T^{00}$ is quadratic in $\pi^0$, we have no reason to expect that it will be properly quantized by \eqref{quantization}. Rather surprisingly, if we press on we will find that in some cases the same procedure that led us to \eqref{pullbackmomentumeigenfunction} leads us to the correct eigenvalue equation for single-particle energy states, too.

More specifically, we begin with the classical total field energy $E =  \int dV \, T^{00} = \int dV \, \left( \frac{1}{2} \delta_{\mu \nu} \pi^\mu \pi^\nu + \frac{1}{2} \frac{m^2 c^2}{\hbar^2} \phi^2 \right)$. We quantize $T^{00} = \frac{1}{2} \delta_{\mu \nu} \pi^\mu \pi^\nu + \frac{1}{2} \frac{m^2 c^2}{\hbar^2} \phi^2$ using \eqref{quantization}, finding

$$ Q(T^{00}) = - \frac{\text{i} \hbar c}{V} \pi^0 \pa{}{\phi} - \frac{1}{2} \pi^0 \pi^0 + \frac{1}{2} \delta_{i j} \pi^i \pi^j + \frac{1}{2} \frac{m^2 c^2}{\hbar^2} \phi^2 $$

We then integrate this over $V$ -- noting again the lack of explicit spacetime dependence -- to find

\begin{equation} \label{totalenergyoperator} Q(E) = \int dV \, Q(T^{0 0}) = - \text{i} \hbar c \pi^0 \pa{}{\phi} - \frac{V}{2} \pi^0 \pi^0 + \frac{V}{2} \delta_{i j} \pi^i \pi^j + \frac{V}{2} \frac{m^2 c^2}{\hbar^2} \phi^2 \end{equation}

If we act this operator on a hypothetical extended phase space state $\Psi = \Psi(x^\mu, \phi, \pi^i)$ we find

$$ Q(E) \Psi = - \text{i} \hbar c \pi^0 \pa{\Psi}{\phi} + \left( - \frac{1}{2} \pi^0 \pi^0 + \frac{1}{2} \delta_{i j} \pi^i \pi^j + \frac{1}{2} \frac{m^2 c^2}{\hbar^2} \phi^2 \right) V \Psi $$

Specializing to the same specific extended phase space state $\Psi_s \propto e^{\frac{1}{A \bar A} \phi \bar \phi}$ that gave us our momentum eigenstate equation now gives us

$$ Q(E) \Psi_s = - \frac{\text{i} \hbar c}{A \bar A} \pi^0 \bar \phi \Psi_s + \left( - \frac{1}{2} \pi^0 \pi^0 + \frac{1}{2} \delta_{i j} \pi^i \pi^j + \frac{1}{2} \frac{m^2 c^2}{\hbar^2} \phi^2 \right) \frac{V}{A \bar A} \Psi_s $$

If we again pullback by the complex-valued energy-momentum eigenfunction $\gamma(x^\mu) = A e^{\text{i} g_{\mu \nu} k^\mu x^\nu}$, we find that

\begin{multline} \label{pullbackenergyeigenfunction} \gamma^* \left( Q(E) \Psi_s \right) = \left( - \frac{\text{i} \hbar c}{A \bar A} \, \text{i} k^0 \gamma \bar \gamma +  \frac{V}{2 A \bar A} \left( g_{\mu \nu} k^\mu k^\nu + \frac{m^2 c^2}{\hbar^2} \right) \gamma^2 \right) \gamma^*(\Psi_s) \\
= \left(\hbar \omega + \frac{m^2 c^2}{\hbar^2} \frac{1}{A \bar A} V \gamma^2 \right) \gamma^*(\Psi_s) \end{multline}

where in the second line we have remembered that $k^0 = \frac{\omega}{c}$ and that $g_{\mu \nu} k^\mu k^\nu = \frac{m^2 c^2}{\hbar^2}$ because $\gamma$ is a solution to the Klein-Gordon equation of motion. 

When $m = 0$, \eqref{pullbackenergyeigenfunction} matches the usual result for the total energy of a single quantum particle in the momentum eigenstate $\vec k$. When $m \neq 0$, $\Psi_s$ is no longer an energy eigenstate, its energy differing from the usual result by the complex-valued, spacetime dependent term $ \frac{m^2 c^2}{\hbar^2} V \frac{A}{\bar A} e^{2 \text{i} g_{\mu \nu} k^\mu x^\mu}$, which is typically infinite.

\section{Discussion} \label{discussion}

The technical results of the previous section may seem somewhat promising to those familiar with canonical quantum field theory, but also confusing in light of the very unusual process that led to them. Before trying to understand the significance of these results, it may be helpful to reiterate the process:

\begin{enumerate}
\item The polysymplectic approach to covariant Hamiltonian field theory gives us a generalized Poisson tensor $\Pi$ and vector-valued polysymplectic potential $\theta$
\item Choosing an arbitrary timelike vector field allows us to closely mimic the standard, finite-dimensional prequantization map of geometric quantization, but on the extended phase space of a polysymplectic Hamiltonian field theory
\item Though none of the resulting operators carries any natural spacetime dependence, and neither the canonical commutation relations nor any particular energy-momentum eigenfunction equations seem to be within reach, the \emph{integrated} commutation relations of canonical quantum field theory can be matched by fixing the parameter $\frac{1}{L^3}$ from the quantization procedure of \eqref{quantization}
\item Careful analysis shows that there is at least one extended phase space wavefunction $\Psi_s \in \mathbb{S}$ such that the natural analog of the total momentum operator -- and, in some cases, the total energy operator -- from canonical quantum field theory yields the standard energy-momentum eigenvalues for single particle wavefunctions upon pullback by single particle energy-momentum eigenfunctions.
\end{enumerate}

As a mathematical enterprise, we might regard this approach as quite successful. Step 1 is mathematically very well founded, not only in flat spacetimes such as those considered here but in arbitrarily curved spacetimes as well. Step 2 requires us to choose an arbitrary timelike vector field (and its dual) to contract with our polysymplectic structures, but such vector fields are guaranteed to exist in globally hyperbolic spacetimes of the type generally considered in quantum field theory on curved spacetimes. In the case of Minkowski space, the integrated commutation relations that result from step 3 are independent of the particular timelike vector field we choose. Through step 4, we see that it is possible to reproduce at least one other (that is, besides the integrated commutation relations) fundamental physical result from canonical quantum field theory through pullback of our geometric structures by appropriate solution sections. Understanding more precisely how best to interpret this last part of the construction beyond single particle energy-momentum eigenstates is an important point to be addressed in future work. 

\section{Conclusions and outlook} \label{conclusions}

Despite its successes, this analysis does not provide an alternative approach to quantum field theory. In particular, nothing has been said about how to find expectation values of operators on states. Though there seem to be several natural candidates for doing this, it is less clear that it is possible to construct any analog of the standard Fock space of canonical quantum field theory, or the unequal time commutation relations, or any of the familiar propagators. Much work remains to show that the alternative approach to quantum fields presented in this analysis can be considered successful even in the case of scalar fields on Minkowski spacetimes.

However, it is my hope that articulating even these few fundamental links between this non-standard approach to the geometric quantization of classical fields and the canonical one may serve to inspire further work in the geometric quantization of classical fields and on exploring the relationship between those programs and the actual physics of quantum field theory.

\section*{Data availability statement}

Data sharing is not applicable to this article as no datasets were generated or analyzed for this analysis.

\section*{Funding} 

Partial funding for this work was received from the Washington and Lee Lenfest Grant program.

\section*{Competing interests}

The author has no competing interests to disclose relating to this analysis.

\bibliographystyle{unsrt}

\bibliography{kspsfps.bib}

\end{document}